\documentclass{article}

     \usepackage[preprint]{neurips_2019}

\usepackage[utf8]{inputenc} 
\usepackage[T1]{fontenc}    
\usepackage{hyperref}       
\usepackage{url}            
\usepackage{booktabs}       
\usepackage{amsfonts}       
\usepackage{nicefrac}       
\usepackage{microtype}      
\usepackage{amsmath}
\usepackage{graphicx}
\usepackage{subfloat}
\usepackage{subcaption}
\graphicspath{ {./figures/} }

\usepackage{amsmath} 
\newcommand{\angstrom}{\textup{\AA}}

\title{Seq-SetNet: Exploring Sequence Sets for Inferring Structures} 

\author{%
 Fusong Ju$^{\text{1,2}}$, Jianwei Zhu$^{\text{1,2}}$, Guozheng Wei$^{\text{1,2}}$, Qi Zhang$^{\text{1,2}}$, Shiwei Sun$^{\text{1,2}}$,   Dongbo Bu$^{\text{1,2}}$\\
  $^{\text{\sf 1}}$Key Lab of Intelligent Information Processing, Institute of Computing Technology, \\Chinese Academy of Sciences, Beijing, 100190, China\\
  $^{\text{\sf 2}}$University of Chinese Academy of Sciences, Beijing, 100049, China\\
  \texttt{\{jufusong, zhujianwei, weiguozheng, zhangqi, dwsun, dbu\}@ict.ac.cn}\\
}

\begin{document}

\maketitle

\begin{abstract}
Sequence set is a widely-used type of data source in a large variety of fields.  A typical example is protein structure prediction, which takes an multiple sequence alignment (MSA) as input and aims to infer structural information from it. Almost all of the existing approaches exploit MSAs in an indirect fashion, i.e., they transform MSAs into position-specific scoring matrices (PSSM) that represent the distribution of amino acid types at each column. PSSM could capture column-wise characteristics of MSA, however, the column-wise characteristics embedded in each individual component sequence were nearly totally neglected.

The drawback of PSSM is rooted in the fact that an MSA is essentially an unordered sequence set rather than a matrix. Specifically, the interchange of any two sequences will not affect the whole MSA. In contrast, the pixels in an image essentially form a matrix since any two rows of pixels cannot be interchanged. Therefore, the traditional deep neural networks designed for image processing cannot be directly applied on sequence sets. 
Here, we proposed a novel deep neural network framework (called Seq-SetNet) for sequence set processing. By employing a {\it symmetric function} module to integrate features calculated from preceding layers, Seq-SetNet are immune to the order of sequences in the input MSA. This advantage enables us to directly and fully exploit MSAs by considering each component protein individually. We evaluated Seq-SetNet by using it to extract structural information from MSA for protein secondary structure prediction. Experimental results on popular benchmark sets suggests that Seq-SetNet outperforms the state-of-the-art approaches by 3.6\% in precision. These results clearly suggest the advantages of Seq-SetNet in sequence set processing and it can be readily used in a wide range of fields, say natural language processing.

\end{abstract}
\section{Introduction}
Sequences usually carry rich structural information.  For example, proteins are linear chains of amino acid residues but always spontaneously fold into specific 3D structural conformations due to interactions among the component amino acids. Similarly, natural language sentences always form complex grammar structures that capture the relationship among the component words. 
These facts together suggest that it is feasible to extract structural information from sequences.  


Inferring protein structure from sequence is extremely important due to the limitations of techniques for protein structure determination. 
Specifically, the widely-used experimental technologies to deduce protein structures, such as X-ray crystallography, NMR spectroscopy, and electron microscopy, have achieved great success; however, these technologies usually cost considerable time and thus cannot keep up with the fast process to acquire protein sequences \citep{berman2000protein, bairoch2005universal}. This limitation emphasizes the importance of predicting protein structure purely from protein sequence. The existing approaches to protein structure prediction can be divided into template-based \citep{roy2010tasser, yang2011improving, ma2012conditional, ma2014mrfalign, wang2016falcon} and {\it ab initio} prediction approaches \citep{simons1997assembly, li2008fragment, xu2012ab}. Accurate prediction of protein structures is highly desired since the functional properties of proteins are largely determined by their three dimensional structures \citep{branden1999introduction}. 



Nearly all approaches to protein structure prediction start with a sequence set. For a target protein, all homologous sequences were first identified and then aligned against the target protein sequence, forming an multiple sequence alignment (MSA).   MSA provide rich evolutionary information of the target protein sequence: each column of an MSA records possible variations of the amino acids at the corresponding position; more importantly, each row of the MSA, i.e., a sequence, describes the correlation among different columns. 

The most popular strategy to exploit MSAs is converting it into position-specific scoring matrices (PSSMs), which simply count the appearance of various types of amino acids at each column. During this converting process, a great deal of information embedded in MSAs are missing, resulting in at least two limitations of PSSMs: 1)  Each component sequence of an MSA carry correlation information among positions; however, PSSMs treat positions of MSA separately and thus cannot describe this correlation information.  2) When MSAs contain only few sequences, the standard deviation of the appearance counts is usually very large, making PSSM not that reliable. Thus, how to directly and effectively exploit MSAs remains a great challenge. 

\begin{figure}[h]
     \centering
    \includegraphics[width=1\textwidth,angle=0]{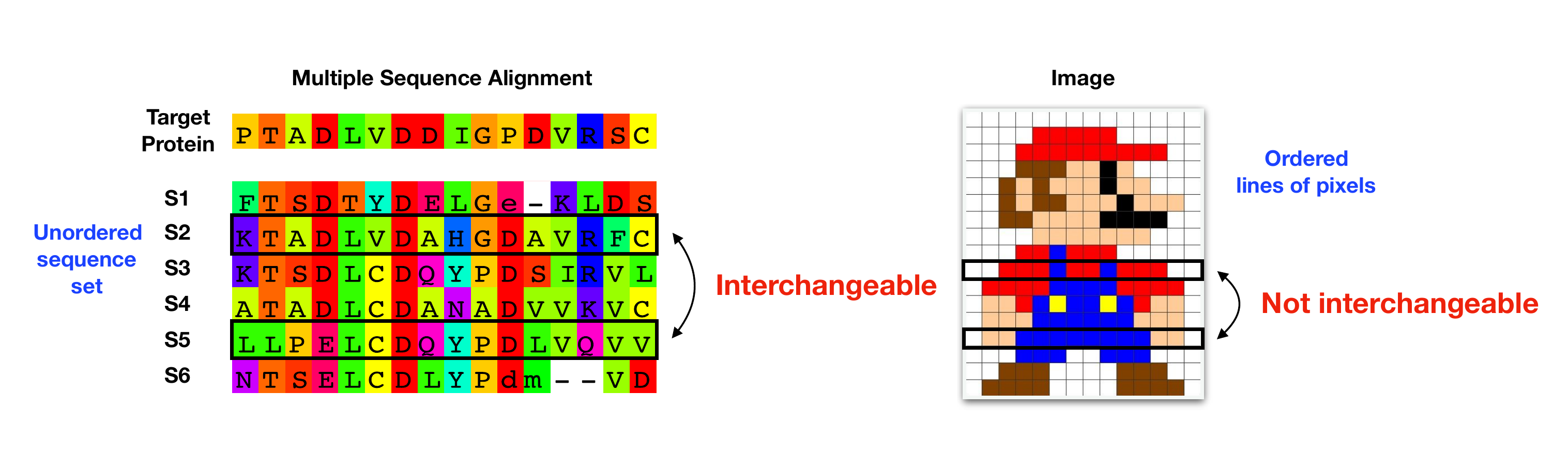}
    \caption{Sequence set essence of MSA versus the matrix essence of images. The sequences in an MSAs from an unordered set, i.e., the interchange of any two sequences will not affect the whole MSA. In contrast, the pixels in an image form a matrix rather than a set: we cannot exchange two rows of pixels of an image with no influence on the image. }
    \label{figure:MSAimage}
\end{figure}

One might suggest to simply borrow the neural network frameworks designed for image processing. However, the sequence set essence of MSAs preclude this strategy. Specifically, the sequences in an MSAs essentially from an unordered set, i.e., the interchange of any two sequences will not affect the whole MSA. In contrast, the pixels in an image essentially form a matrix rather than a set; thus, we cannot exchange two rows of pixels of an image with no influence on the image (Fig. \ref{figure:MSAimage}). Consequently, the deep neural networks designed for image processing cannot be directly applied on MSAs, and a framework specially designed for sequence sets is highly desired.

In this work, we present such a neural network framework (called Seq-SetNet) suitable for sequence set processing. The key module of Seq-SetNet is a symmetric function: this function aims to integrate features calculated from preceding layers but will not be influenced by order of the input sequences. We applied  Seq-SetNet to predict protein secondary structures. On popular benchmark sets, our approach outperformed the state-of-the-art approaches by 3.6\% in precision. This result clearly suggest the advantage of Seq-SetNet in processing sequence sets. It is worth pointed out that our approach could be readily applied to other sequence set processing task such as natural language processing.

\section{Methods}
\begin{figure}[h]
  \begin{subfigure}{.55\linewidth}
    \centering
    \includegraphics[width=1\textwidth]{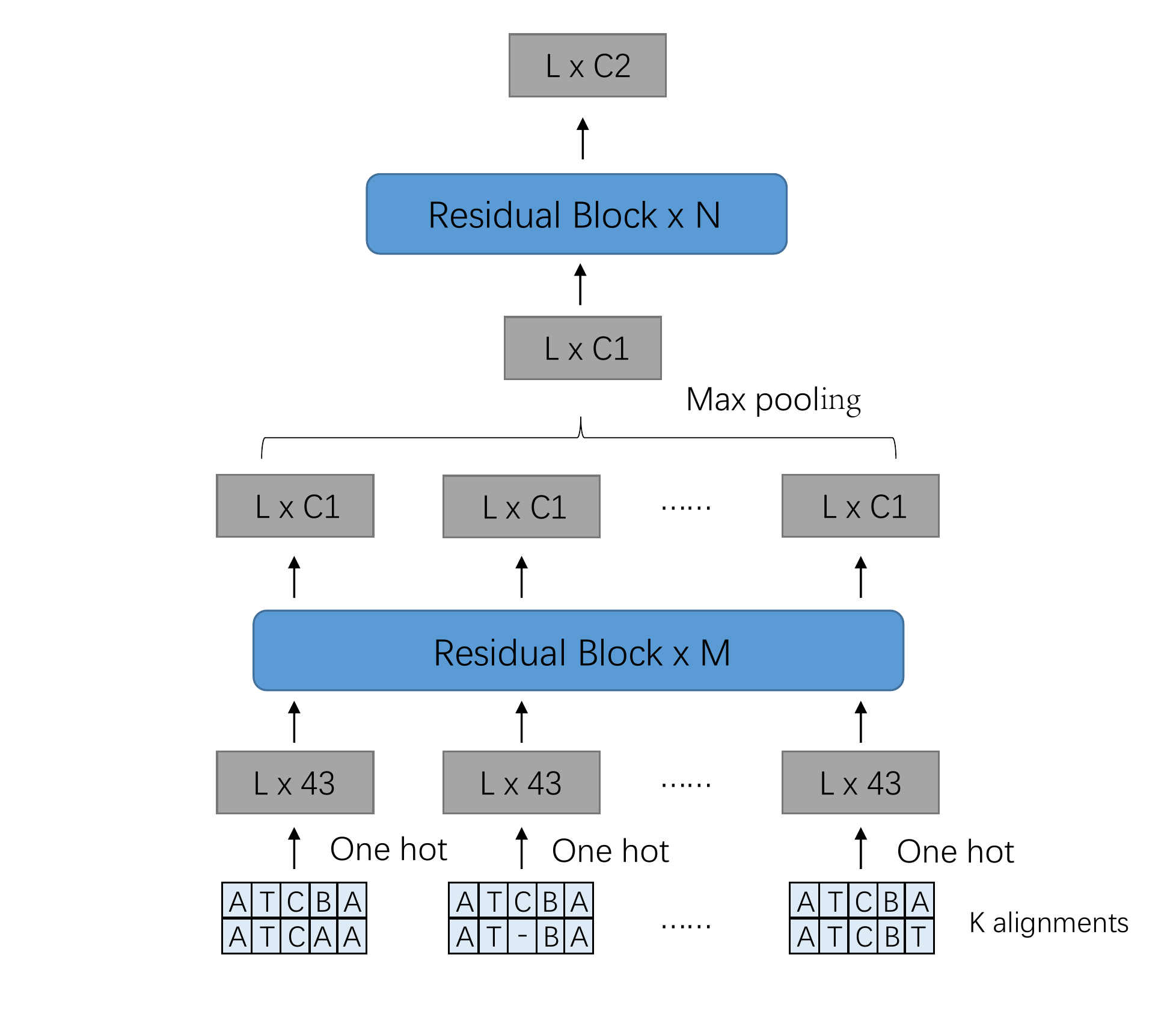}
    \caption{Seq-SetNet architecture}
    \label{figure:0b}
  \end{subfigure}
  \begin{subfigure}{.45\textwidth}
    \centering \includegraphics[width=.89\textwidth]{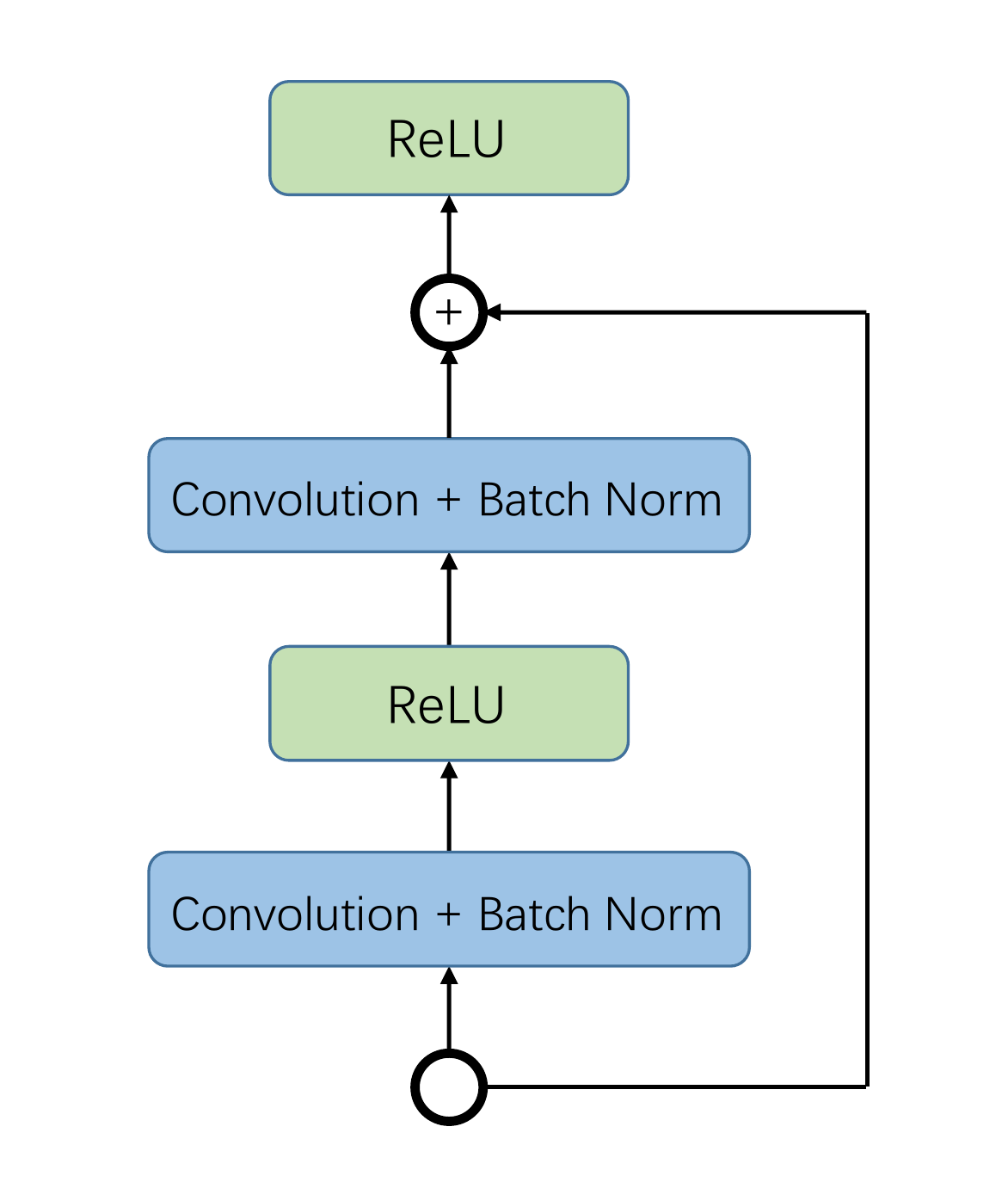}
    \caption{Residual Block}
    \label{figure:0a}
  \end{subfigure}
  \caption{Seq-SetNet architecture and the details of residual blocks used in Seq-SetNet }
  \label{figure:0}
\end{figure}

The architecture of Seq-SetNet is illustrated in Figure \ref{figure:0}. Specifically, Seq-SetNet consists of two key modules: stacked residual blocks for feature extraction, and the max
pooling layer as a symmetric function to aggregate features calculated by preceding layers. 
Given a MSA containing $K$ sequence similar to a target protein, we first converted each component sequence into a sequence pair: each sequence pair was constructed through aligning a component sequence against the target protein sequence. Thus we obtained a total of $K$ sequence paris. Seq-SetNet takes these $K$ sequence pairs as input and works as follows: 
\begin{enumerate}
\item  First, Seq-SetNet encodes each sequence pair into $L \times 43$ initial features using the one-hot technique. Here $L$ denotes the length of the target protein sequence.
\item  Next, Seq-SetNet extracts position-correlation features using $M$ stacked residual blocks, and then 
aggregates the extracted features using max pooling. 
\item  Finally, Seq-SetNet uses $N$ residual blocks to extract structural information as final output. 
\end{enumerate}

The underlying rationale of our design are described in more details as below. 

\subsection{Stacked residual blocks for extracting position-correlation within individual sequence}
Previous studies have shown that for proteins, each residue is tightly correlated with its neighboring residues. The local interaction among neighboring residues leads to considerable preference for local structural conformations 
\citep{betancourt2004local,keskin2004relationships,jha2005helix}. To capture the correlation among positions, we introduced convolution layers, which contain multiple filters of length 5 to calculate correlation within 5 neighboring residues. As shown in Figure \ref{figure:0b}, we used residual block consisting of 2 convolution layers, 2
batch normalization layers \citep{ioffe2015batch} and 2 activation layers. 

Beside local interactions among neighboring residues, long-range interactions play important roles in putting local structures in appropriate positions. To account the long-range interactions, we used stacked residual blocks together with ResNet technique (Fig. \ref{figure:0a}). This way, Seq-SetNet works in a hierarchical fashion and thus integrate both local correlation and long range correlations as well \citep{srivastava2015training,szegedy2015going,he2016deep}. 

It is worth pointed out that Seq-SetNet processes each sequence pair individually; thus, the correlation information carried by this sequence pair can be exploited effectively. In contrast, PSSM considers each position separately and therefore totally neglects the correlation among positions.


\subsection{Symmetric function for removing the influence of sequence order} 
Seq-SetNet processes the $K$ sequence pairs one by one. To make Seq-SetNet immune to the oder of the input sequence pairs, we employed a symmetric function to process the calculated correlation features.

The symmetric function $f : \{\mathbb{R}^{L}, \ldots \mathbb{R}^{L}\} \rightarrow \mathbb{R}^{C_1}$ is defined on a sequence set and calculated as below:
\[
f(x_{1}, \ldots, x_{K}) = h_2 \circ g(h_1(x_{1}), \ldots, h_1(x_{K}))
\]
Here, $x_{i}$ represents the $i$-th sequence in MSA, $h_1 : \mathbb{R}^{L} \rightarrow \mathbb{R}^{C_1}$  represents  the stacked residual blocks for extracting correlation features for 
individual sequence pairs,  $g : \{\mathbb{R}^{C_1}, \ldots \mathbb{R}^{C_1}\} \rightarrow \mathbb{R}^{C_1}$  represents a max pooling function as symmetric function, and $h_2 : \mathbb{R}^{C_1} \rightarrow \mathbb{R}^{C_2}$ denotes 
another stacked residual blocks to extract global feature from the whole MSA.
The max pooling operator $g(h_1(x_{1}), \ldots, h_1(x_{K}))$ reports the maximum among the input $K$ sequences $x_{1}, \ldots, x_{K}$, therefore immune to order of these input sequences. 

One might suggest to augment the input sequences by enumerating all possible permutations of the sequences and feeding these permutations to a RNN model. Using this data-augmentation strategy, the output will be independent of the order of the input sequences. However, previous study has shown that order cannot be totally omitted \citep{vinyals2015order}. More importantly, the number of possible permutations grows exponentially to the number of input sequences, making this strategy inapplicable to MSA containing hundreds of proteins.
%
%

\section{Experiment setting}
In this section, we show how to apply Seq-SetNet to predict protein secondary structure. The experiment settings are described as follows. 

\textbf{Dataset.} We used a large non-homologous sequence
and structure dataset with 6128 proteins (after filtering out redundant proteins), and divided it
randomly into training (5600 proteins), validation (256 proteins), and testing (272 proteins) sets. This
dataset was produced with PISCES Cull PDB server \citep{wang2003pisces} which is
commonly used for evaluating structure prediction algorithms. We retrieved a
subset of solved protein structures with better than 2.5$\angstrom$ resolution while
sharing less than 30\% identity (widely-used criteria in previous studies, say 
\citep{wang2010protein}).

\textbf{Benchmark.} We evaluated the accuracy of predicted secondary structure. To make evaluation more objective, we performed a separate evaluation on CB513 dataset, which is independent with the training set. Again, a filtering operation was performed to guarantee no overlap between CB513 dataset and the training set (sequence identity < 25\%). 

\textbf{Feature.} We used the input features described in \citep{zhou2014deep},
but replaced position-specific scoring matrix(PSSM) with raw multiple sequence
alignment (MSA). In particular, for each protein sequence, we ran PSI-BLAST
\citep{altschul1997gapped} with E-value threshold 0.001 and 3 iterations to
search UniRef90\citep{uniprot2007universal} to acquire MSA. We used a binary
vector of 21 elements to indicate the amino acid type at position i of input
protein sequence (20 for amino acid types, 1 for unknown type).  For each
alignment, we also encoded each position with 22-length one-hot vector (20 for
amino acid types, 1 for unknown type, 1 for gap type).

\textbf{Label.} Here, we adopted the 8-states secondary structure labels. For each protein with known structure, we executed DSSP  \citep{kabsch1983dictionary} to generate 8-state secondary structure label (denoted as SS8) for each residue. 

\textbf{Classification layer.}  used two fully connected layers as the final layers, which generate marginal probability of the 8 labels.

\textbf{Loss.} We trained Seq-SetNet using cross-entropy loss function, and aims to maximize the probability of reporting correct secondary structure labels.


\section{Results and discussion}
\subsection{Accuracy of secondary structure prediction}
Table \ref{table:1} shows the performance comparison between Seq-SetNet
and four of the best existing methods, including SSPro\citep{cheng2005scratch},
ICML2014\citep{zhou2014deep}, DeepCNF-SS\citep{wang2016protein}, and
MUFOLD\citep{fang2018mufold}. 
 All these methods have identical training dataset and thus the comparison is fair. Besides, SSPro was run without using template information. As shown in this table, Seq-SetNet obtained the highest
accuracy (74.2\%), 3.6\% higher than state-of-the-art
methods.

\begin{table}[h!]
  \begin{center}
    \begin{tabular}{cc} 
      \toprule
      Tool & Prediction accuracy\\
      \midrule
      SSPro & 0.636 \\ 
      ICML2014 & 0.664 \\
      DeepCNF-SS & 0.683 \\
      MUFOLD-SS & 0.706 \\
      \textbf{Seq-SetNet} & \textbf{0.742} \\
      \bottomrule
    \end{tabular}
  \end{center}
  \caption{Accuracy of Seq-SetNet and four existing approaches to prediction of protein secondary structure. Dataset: CB513}
  \label{table:1}
\end{table}

We further examined the prediction accuracy for each of the 8 labels. As shown in Table \ref{table:2},  for four major states H, E, L and T, Seq-SetNet achieved extremely high accuracy above 80\%. However, for states G and S, the prediction accuracy is not that high. A reasonable explanation is that states H, E, L and T are popular and we have sufficient residues with these states in the training set. In contrast, the states G and S are not popular. This 
unbalanced label problem could be circumvented using the weighting strategy, which is left as one of the future works. 

\begin{table}[h!]
  \begin{center}
    \begin{tabular}{ccccr}
      \toprule
      SS8 label & Precision & Recall & F1-score & Descriptions\\
      \midrule
      B & 0.5534 & 0.0488 & 0.0898 & $\beta$-bridge\\
      E & 0.8006 & 0.8498 & 0.8245 & $\beta$-strand\\
      G & 0.5444 & 0.4162 & 0.4717 & $3_{10}$-helix\\
      H & 0.8871 & 0.9405 & 0.9130 & $\alpha$-helix\\
      L & 0.5991 & 0.7360 & 0.6601 & loop or irregular\\
      S & 0.6241 & 0.3058 & 0.4105 & bend\\
      T & 0.6316 & 0.5926 & 0.6115 & $\beta$-turn\\
      I & - & - & - & $\pi$-helix\\
      \bottomrule
    \end{tabular}
  \end{center}
  \caption{Prediction accuracy of Seq-SetNet for each of the SS8 states. Dataset: CB513}
  \label{table:2}
\end{table}

\subsection{Protein secondary structure prediction using Seq-SetNet: a case study} 
\begin{figure}[h]
  \centering
  \includegraphics[width=1\textwidth]{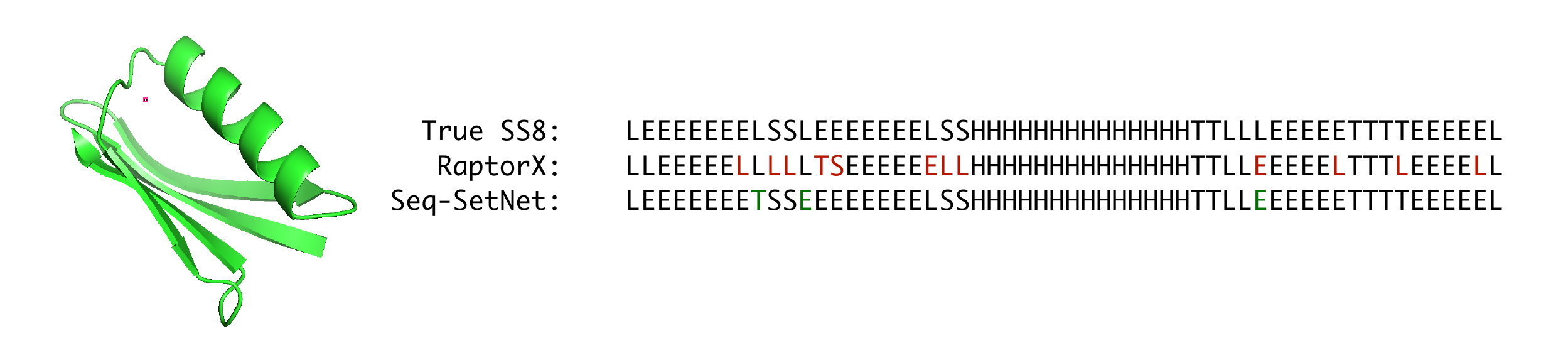}
  \caption{Predicted secondary structure for protein {\tt 1PBG}  using RaptorX and  Seq-SetNet. Here the amino acids in red and green indicates the incorrect prediction by RaptorX and Seq-SetNet, respectively. }
\label{casestudy}
\end{figure}

\begin{figure}[h]
  \centering
  \includegraphics[width=1\textwidth]{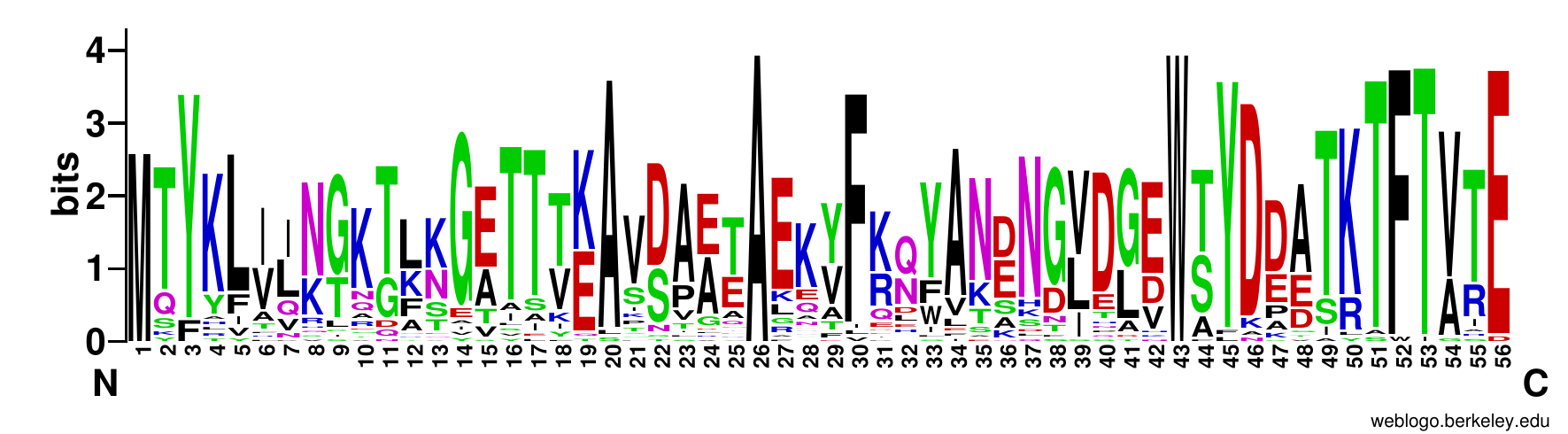}
  \caption{Multiple sequence alignment of protein {\tt 1PBG} (shown in web logo format). Here the $x$ axis represents  position of the protein, and $y$ axis indicates the sequence conservation at that position. The height of symbols within the stack indicates the relative frequency of each amino acid type. }
\label{logo}
\end{figure}

We further examined the prediction results for protein  {\tt 1PBG}.  As shown in Figure \ref{casestudy}, RaptorX reported incorrect prediction for 12 amino acids whereas Seq-SetNet committed errors at only 3 amino acids (including 10T, 13E, and 41E). The reason for the failure of RaptorX is that for this protein, only 36 homologous proteins were identified and used to construct MSA (Fig. \ref{logo}). Using so few homologous protein sequences, it is unreliable to calculate the distribution over a total of 20 amino acids types.   

\subsection{Analysis of the contribution by each component sequence}
To investigate whether Seq-SetNet could exploit all component sequences in MSA, we 
investigated the contribution by each component sequence.   For this aim, we calculated the output of the max
pooling layer for protein {\tt 1PBG}. The output is a $L\times C_1$ tensor (denoted as $F$), where
$L=56$ indicates protein length, and $C_1=40$ indicates 
channel size of the last convolution layer before the max-pooling
layer. 

\begin{figure}[h]
  \centering
  \includegraphics[width=0.7\textwidth]{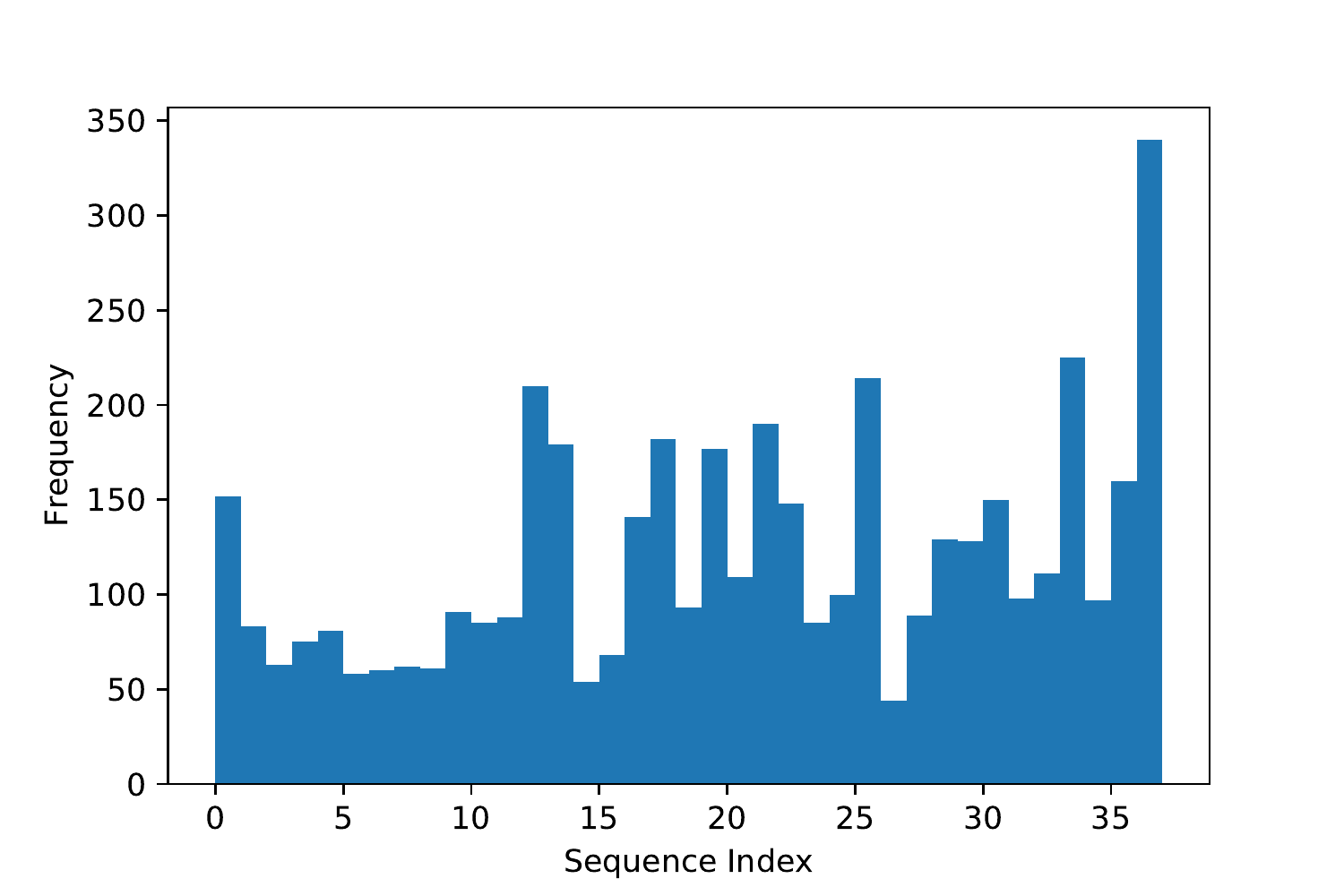}
  \caption{Frequency of component sequences to be selected as maximum. Here, the $x$ axis indicates the sequence indices, while the $y$ axis indicates the frequency for a sequence to be selected as maximum by the max pooling operation. Protein: {\tt 1PBG}. }
\label{contribution}
\end{figure}

We calculated the contribution by each of the 36 component protein, i.e., the frequency of each sequence was selected as the maximum by the max pooling operation. As shown in Figure \ref{contribution}, each component sequence has the chance to be selected as maximum. However, the frequencies of these sequences differed greatly. For example, the 36th sequence was selected as maximum for over 300 times whereas the 27th sequence was selected for only 47 times. This clearly suggests that some sequences in MSA are more important than others. However, the PSSM technique treats each sequence equally when counting the appearance of 20 amino acid types, therefore inevitably leading to significant variations.

%


\subsection{Relationship between prediction accuracy and the number of component sequences in MSA }

It is well known that the accuracy of protein structure prediction is highly related to the number of homologous proteins of the target protein. The more homologous proteins were identified, the more possibility to acquire reliable prediction. On the other side, when only few homologous proteins were identified, it is difficult for the existing approaches to make reliable prediction. 

\begin{figure}[h]
  \centering
  \includegraphics[width=0.7\textwidth]{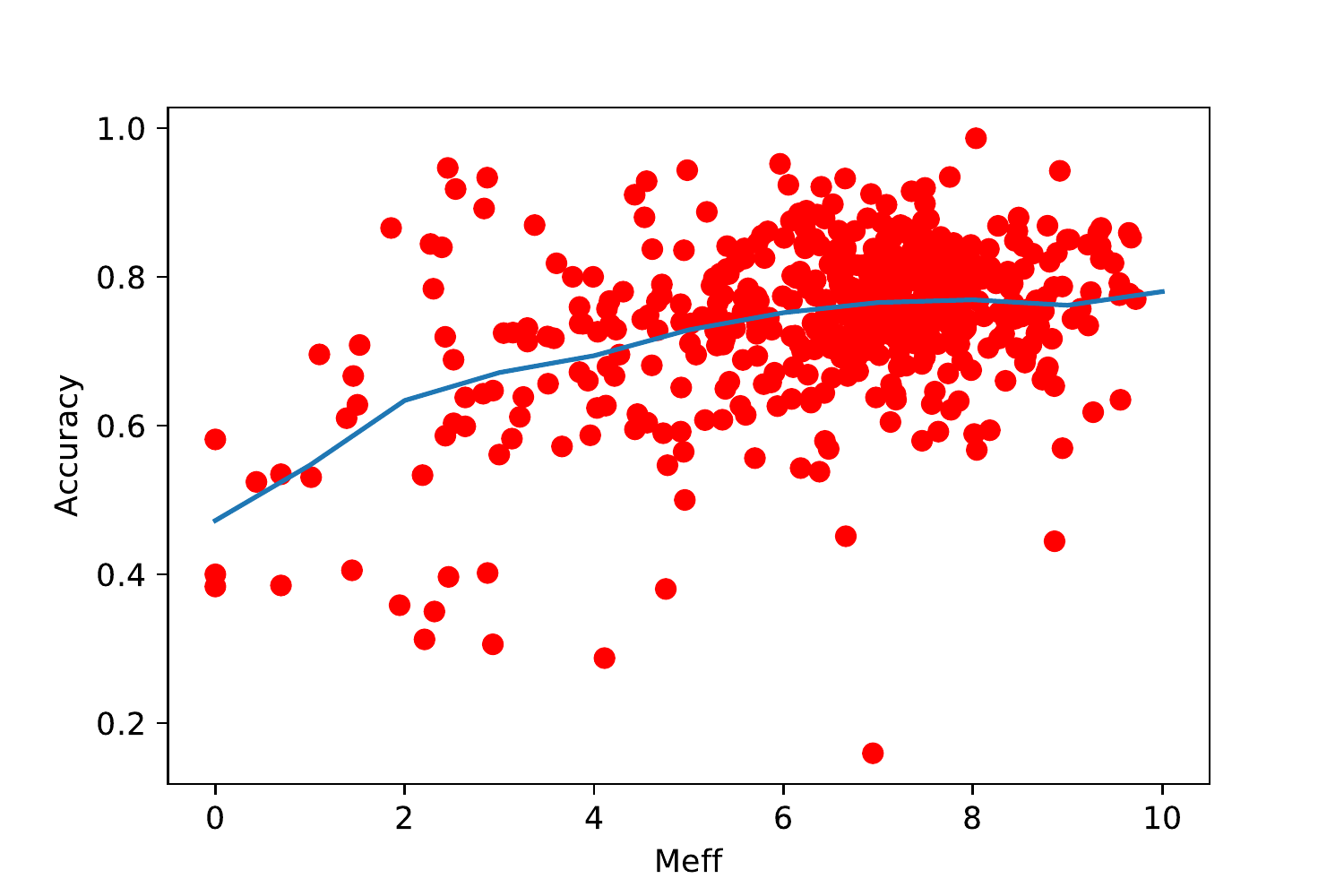}
  \caption{Relationship between prediction accuracy and the number of component sequences in MSA (measured using Meff) }
\label{meff}
\end{figure}

Here we further investigated the power of Seq-SetNet when only limited number of homologous proteins are available. As shown in Figure \ref{meff}, when the effective number of homologous protein (denoted as Meff) is over 4, Seq-SetNet can achieve prediction accuracy over 0.70. The prediction accuracy increases slowly with Meff,  suggesting that Seq-SetNet can apply for target protein with only a few homologous proteins. This advantage is rooted in that Seq-SetNet processes each component sequence individually and thus exploits full information embedded in MSA. 

\section{Conclusion}
In this work, we present a novel deep neural network Seq-SetNet for sequence set processing. 
Using Seq-SetNet, we can 
extracts structural information directly from MSA rather than
compressing them into position-specific scoring matrix. Furthermore, Seq-SetNet does not need any other
manually designed or otherwise hand-picked extra feature as input; instead, Seq-SetNet exploits the capability that deep networks could 
automatically extract relevant feature from the raw data. The experimental results clearly suggest the advantages of Seq-SetNet in sequence set processing and it can be readily used in a wide range of fields, say natural language processing.

%


\bibliography{ply}
\end{document}